\documentclass[aps,reprint,showpacs,twocolumn,superscriptaddress,longbibliography,floatfix]{revtex4-2}
\usepackage[colorlinks,linkcolor=blue,citecolor=blue,anchorcolor=blue,urlcolor=blue]{hyperref}
\usepackage{enumitem}
\usepackage{bm,amssymb,amsmath}
\usepackage{dcolumn}
\usepackage{multirow}
\usepackage{graphicx}
\usepackage{xcolor}
\usepackage[normalem]{ulem}

\begin{document}

\title{Equivalent-neighbor $k$-core percolation in two dimensions}

\date{\today}

\author{Qiyuan Shi}
\affiliation{School of Physics, Hefei University of Technology, Hefei, Anhui 230009, China}
\affiliation{Department of Modern Physics, University of Science and Technology of China, Hefei, Anhui 230026, China}

\author{Ming Li}
\email{lim@hfut.edu.cn}
\affiliation{School of Physics, Hefei University of Technology, Hefei, Anhui 230009, China}

\author{Youjin Deng}
\email{yjdeng@ustc.edu.cn}
\affiliation{Department of Modern Physics, University of Science and Technology of China, Hefei, Anhui 230026, China}
\affiliation{Hefei National Research Center for Physical Sciences at the Microscale, University of Science and Technology of China, Hefei, Anhui 230026, China}
\affiliation{Hefei National Laboratory, University of Science and Technology of China, Hefei, Anhui 230088, China}

\begin{abstract}
We perform large-scale numerical simulations to investigate the critical behavior of $k$-core percolation in two
dimensions with an extended interaction range $r$. By systematically varying both the core index $k$ and the interaction
range $r$, we construct a comprehensive phase diagram in the $(k,r)$ plane. In contrast to $k$-core percolation in
infinite dimensions, no hybrid transition is observed in two dimensions: the phase diagram contains only a continuous
transition regime and a strictly first-order regime, separated by a tricritical or critical-end point $(k_s,r_s)$. For
$k<k_s$ and $r<r_s$, the transition is continuous and belongs to the universality class of standard two-dimensional (2D)
percolation. For $k>k_s$ and finite $r>r_s$, the transition is discontinuous, with no hybrid features or critical
singularities. In this first-order regime, the pseudocritical point approaches the critical point as $1/\ln L$, where
$L$ is the linear system size, distinct from the $L^{-d}$ scaling typical of conventional thermodynamic first-order
transitions in $d$ dimensions. This logarithmic finite-size drift is consistent with a nucleation-driven mechanism, in
which rare voids trigger the collapse of the finite-range $k$-core. These results demonstrate that geometric constraints
can fundamentally alter the nature of $k$-core percolation found in finite dimensions.
\end{abstract}

\maketitle

\section{Introduction} \label{sec-intro}

Phase transitions and critical phenomena are central topics in statistical physics, providing a
unifying framework for understanding collective behavior in interacting many-body systems.
Traditionally, phase transitions are classified into continuous and
discontinuous (first-order) types~\cite{Ma2018}. In continuous transitions,
the order parameter varies smoothly across the critical point, accompanied by a
diverging correlation length, scale invariance, and universal power-law behavior.
In contrast, discontinuous transitions are characterized by a finite jump of the
order parameter, latent heat, and a finite correlation length, without critical fluctuations.

Beyond this classical dichotomy, a remarkable class of transitions has been identified
in which a discontinuous jump of the order parameter coexists with diverging fluctuations
and critical scaling laws. Such transitions, commonly termed hybrid phase transitions,
have attracted growing interest due to their unconventional nature and broad relevance
in jamming systems~\cite{Sellitto2005,Schwarz2006,Toninelli2006}, kinetically constrained
systems~\cite{Sausset2009,Sellitto2015}, Ising-like models~\cite{Bonamassa2025,Korbel2025},
and networked systems~\cite{Dorogovtsev2006,Goltsev2006,Buldyrev2010,Lee2016a,Gross2022,Cho2026}.

A paradigmatic and minimal framework for investigating hybrid phase transitions and their
crossover to continuous criticality is provided by percolation theory. In standard
percolation~\cite{Stauffer1992}, one studies the connectivity properties of a system
after a fraction of sites or bonds is randomly removed, where the emergence of a giant
connected cluster is associated with a continuous phase transition. Generalized percolation
models incorporating cooperative effects or dynamical constraints can, however, exhibit
hybrid critical behavior. A prominent example is the $k$-core percolation
model~\cite{Chalupa1979,Seidman1983,Dorogovtsev2006,DeGregorio2016}, in which
a site remains active only if it is connected to at least $k$ other active sites.
This local stability constraint induces an iterative pruning process, generating
cascades of site removals and collective collapse phenomena.

The $k$-core model has been widely employed to characterize the structural organization
and robustness of complex networks~\cite{Kong2019}, and as a minimal model for jamming
and dynamical arrest~\cite{Sellitto2005,Schwarz2006}. Historically, this class of models
was often referred to as bootstrap percolation~\cite{Chalupa1979,Kogut1981,Adler1991,Branco1993}.
In modern terminology, however, bootstrap percolation typically denotes an activation process,
conceptually distinct from the pruning dynamics of $k$-core
percolation~\cite{Baxter2010,Baxter2011,DiMuro2019,DiMuro2020}.

In mean-field settings, such as on Bethe lattices and complete graphs (CGs), the critical
behavior of $k$-core percolation is now well understood~\cite{Chalupa1979,Pittel1996,Dorogovtsev2006,Goltsev2006}.
For $k=1$, the model reduces to standard percolation and exhibits a continuous
transition with order-parameter exponent $\beta=1$. For $k=2$, the transition remains
continuous and occurs at the same percolation threshold, but with a different
order-parameter exponent $\beta=2$. In contrast, for $k \ge 3$, the system undergoes
a hybrid phase transition, in which the order parameter--defined as the relative size
of the $k$-core--exhibits a discontinuous jump at criticality, while the susceptibility
diverges and power-law scaling emerges. This mixed character, combining features of both
continuous and discontinuous transitions, has been confirmed by extensive numerical
simulations on Erd\H{o}s-R\'enyi (ER) graphs~\cite{Lee2016}.

This hybrid phenomenology is generally attributed to the nonlocal effects induced by
cascade pruning~\cite{Goltsev2006,Farrow2007,Iwata2009,Gao2024}. In locally tree-like
structures, such as Bethe lattices or ER graphs, pruning avalanches can propagate over
arbitrarily long distances without being intercepted by short loops, leading to a unique
interplay between global cascades and critical fluctuations. We note that, on strictly
tree-like graphs, finite $k$-cores with $k \ge 2$ cannot exist in the thermodynamic limit.
Furthermore, even in systems containing loops, the existence of a nontrivial $k$-core depends
sensitively on the coordination number of the lattice. On $d$-dimensional hypercubic lattices,
when $k>d$, no extensive $k$-core survives as soon as a nonzero fraction of sites is
removed~\cite{Schonmann1990,Schonmann1992}.

In finite-dimensional systems, however, it remains unclear whether this hybrid character survives
or is replaced by a first-order transition. For $d$-dimensional regular lattices, analytical
expansions in powers of $1/d$ suggest that weak finite-dimensional corrections do not immediately
destroy the mixed nature of the transition, implying that hybrid behavior may persist in
sufficiently high, yet finite, dimensions~\cite{Harris2005}. In contrast, the $M$-layer
construction indicates that the mean-field hybrid transition does not survive in physical
dimensions~\cite{Rizzo2019}.

For $k=2$, field-theoretic analyses show that, unlike the mean-field case, $k$-core
percolation belongs to the universality class of standard percolation~\cite{Harris1983,Harris1983a}.
Even for $k=3$ on the three-dimensional cubic lattice, numerical simulations suggest that the
transition remains continuous, with critical exponents consistent with those of standard
three-dimensional percolation~\cite{Branco1999}. On four-dimensional hypercubic lattices,
numerical evidence indicates that $k=4$ exhibits a first-order transition,
whereas the $k=3$ case remains continuous~\cite{Parisi2008}, although its universality
class was not firmly established. Recently, it has been suggested that in two dimensions,
continuous, first-order, and hybrid-like transitions may all occur and even
cross over by tuning the coordination number of a 2D lattice~\cite{Xue2024}.

Taken together, the existing results remain inconclusive regarding how finite-dimensional systems
connect to the infinite-dimension limit, where hybrid transitions are firmly established
both theoretically and numerically. To bridge this gap, it is natural to consider
$k$-core percolation on equivalent-neighbor
lattices~\cite{Luijten1996,Luijten1997,Qian2016,Ouyang2018}, in which each site
interacts with all neighbors within a finite range $r$. For $r\to\infty$,
the lattice becomes equivalent to a CG and exhibits mean-field behavior.

It is instructive to first recall the cases $k=1$~\cite{Ouyang2018} and $k=2$~\cite{Harris1983,Harris1983a}, for which
the transition is continuous. For both cases, as long as the interaction range $r$ is finite, regardless of how large it
is, the system remains genuinely $d$-dimensional, and renormalization-group theory predicts that the transition belongs
to the standard $d$-dimensional percolation universality class. Thus, a qualitative crossover to mean-field criticality
should occur only in the CG limit $r\to\infty$.

Interestingly, this crossover is more subtle for $k=2$ than for $k=1$. For standard percolation ($k=1$), the critical
properties in sufficiently high dimensions (above the upper critical dimensionality, $d > d_u=6$) already coincide with
those on CG. For example, the largest-cluster size scales as $C_1 \sim V^{d_f}$, where $V$ is the system volume, and
$d_f=2/3$ is the volume-fractal dimension. The Fisher exponents are both $\tau=5/2$ for the cluster-size distribution
that scales as $n(s) \sim s^{-\tau}$~\cite{Stauffer1992}. In contrast, for $k=2$, corresponding to leaf-free
clusters~\cite{Huang2018}, the critical behavior in finite dimensions remains fundamentally different from the CG limit.
The volume-fractal dimensions are $d_f(\mathrm{CG})=1/3$ and $d_f(d>6)=2/3$, and the Fisher exponents are
$\tau(\mathrm{CG})=1$ and $\tau(d>6)=5/2$~\cite{Huang2018}. Namely, for $d>6$, the qualitative change is absent for
$k=1$ but present for $k=2$. This is because critical percolation clusters on the CG are mostly tree-like. After the
deletion of leaves, the total number of sites in the remaining $2$-core clusters scales as $\ln V$ rather than $V$.

These observations naturally raise the question of what happens for $k>2$, where the local constraint becomes even more
stringent. As the interaction range is tuned from the CG limit ($r \to \infty$) to large but finite values of $r$, two
distinct scenarios may arise. On the one hand, a hybrid transition could persist at finite $r$, potentially giving rise
to a new universality class beyond standard percolation. On the other hand, the critical singularity may disappear
altogether, leaving a first-order transition. Distinguishing between these possibilities is one of the central goals of
this work.

We address these possibilities by systematically investigating $k$-core percolation on 2D triangular 
lattices with a tunable interaction range $r$. By varying $r$, we control the effective coordination number 
$z(r)$ and follow the crossover from the short-range 2D regime toward the CG limit. For $k \ge 3$, where 
the transition on CGs is genuinely hybrid, our large-scale Monte Carlo simulations find no evidence of a 
hybrid transition at any finite $r$ in two dimensions. Instead, for small $k$ or short interaction 
range $r$, the system exhibits a continuous transition belonging to the standard 2D percolation 
universality class. Increasing either $k$ or $r$ drives the system into the first-order regime, 
while the hybrid transition survives only at the singular limit $r\to\infty$. The boundary $k_s(r)$ 
separating the continuous and discontinuous regimes decreases with increasing $r$ and approaches 
$k_s=2$ for large $r$. In the first-order regime, the finite-size shift of the transition point 
follows the scaling form $1/\ln L$ ($L$ being the linear system size), in quantitative agreement 
with the nucleation mechanism predicted for bootstrap percolation~\cite{Cerf2002,Holroyd2003}. 
By contrast, in infinite-dimensional systems the critical fluctuations exhibit power-law scaling 
with the system volume~\cite{Gao2024}. The absence of such collective critical fluctuations in 
finite dimensions indicates that 2D $k$-core percolation is governed by rare localized void-nucleation 
events, rather than by extended critical avalanches.

The remainder of this paper is organized as follows. In Sec.~\ref{sec-model}, we describe the
simulation algorithm and measured observables. In Sec.~\ref{sec-cg}, we present numerical results
for CG as a benchmark. Our main results for 2D lattices, including universality analysis and the
characterization of the discontinuous transition, are reported in Sec.~\ref{sec-2d}.
The phase diagram of the $k$-core percolation transition, along with related models,
is analyzed from a renormalization-group flow perspective in Sec.~\ref{sec-rfi}.
Finally, we conclude with a summary and discussion in Sec.~\ref{sec-con}.

\section{Models, Algorithms, and Observables} \label{sec-model}

\subsection{Models}

The $k$-core of a given graph or lattice is defined as the maximal subgraph in which every site has
at least $k$ neighbors within the same subgraph. Starting from a bond percolation configuration,
where each bond is independently occupied with probability $p$, the $k$-core is obtained through
an iterative pruning process. In this process, all sites with fewer than $k$ neighbors are recursively removed
together with their incident bonds, until no such sites remain. The remaining cluster constitutes
the $k$-core corresponding to the occupation probability $p$.

We focus primarily on triangular lattices of linear size $L$ with periodic boundary conditions,
containing $V=L^2$ sites. To tune the effective coordination number, we introduce an interaction
range $r$ by connecting each site to all sites within a Euclidean distance $r$. The original
triangular lattice corresponds to $r=1$, for which each site has coordination number $z=6$.
In general, the coordination number $z(r)$ scales as $\sim r^2$ for large $r$.
For reference, Table~\ref{tab1} lists the coordination number $z(r)$ for several interaction ranges used below.
For any finite $r$, the system remains strictly 2D in the thermodynamic limit $L\to\infty$. This
construction therefore enables a systematic investigation of connectivity effects on the
$k$-core percolation transition in two dimensions.

As a mean-field reference, we also study $k$-core percolation on CG, corresponding to $r\to\infty$.
Applying a bond occupation probability $p$ to a complete graph of $V$ sites generates a random
graph with an average number of bonds $pV(V-1)/2$, which is equivalent to an ER random graph with
average degree (average coordination number) $q = p(V-1)$. It is therefore convenient to use the average degree
$q$ as the control parameter. In the thermodynamic limit,
the critical values are $q_c=1$ for $k=1$ and $k=2$~\cite{Chalupa1979,Pittel1996,Dorogovtsev2006,Goltsev2006}.

\begin{table}[b]
\centering
\caption{Coordination number $z(r)$ for representative interaction ranges on the triangular lattice.}
\label{tab1}
\begin{ruledtabular}
\begin{tabular}{ccccccccc}
$r$ & 1 & 2 & 3 & 4 & 5 & 6 & 16 & 64\\
\hline
$z(r)$ & 6 & 18 & 36 & 60 & 90 & 126 & 930 & 14844
\end{tabular}
\end{ruledtabular}
\end{table}

For $k \ge 3$, the mean-field transition is determined by the local minimum of the
function~\cite{Pittel1996}
\begin{equation}
f_k(\lambda)=\frac{\lambda}{\sum_{i=k-1}^{\infty}\frac{e^{-\lambda}\lambda^i}{i!}},
\qquad \lambda>0 .
\end{equation}
Let $\lambda_c$ denote the position of this local minimum, satisfying
$f_k'(\lambda_c)=0$ and $f_k''(\lambda_c)>0$. The critical average degree is
\begin{equation}
q_c=f_k(\lambda_c).
\end{equation}
At criticality, the fraction of sites in the final giant $k$-core (see
Sec.~\ref{sec-obs}) is denoted by $m_0$ and is given by
\begin{equation}
m_0=\sum_{i=k}^{\infty}\frac{e^{-\lambda_c}\lambda_c^i}{i!}.
\end{equation}
Numerical minimization gives the values listed in Table~\ref{tab2}.

\begin{table}
\centering
\caption{Mean-field thresholds and jump sizes for $k$-core percolation on CGs.}
\label{tab2}
\begin{ruledtabular}
\begin{tabular}{ccccc}
$k$ & 3 & 4 & 5 & 6\\
\hline
$q_c$ & 3.3509 & 5.1494 & 6.7992 & 8.3653\\
$m_0$ & 0.2676 & 0.4332 & 0.5492 & 0.6329
\end{tabular}
\end{ruledtabular}
\end{table}

\subsection{Algorithms}

The identification of the $k$-core for a given bond configuration is performed using a standard
recursive pruning procedure, which is computationally efficient and does not constitute a bottleneck
in the simulation. The main computational challenge instead lies in generating bond configurations
when the coordination number is large and the critical occupation probability $p_c$ is small, as in
extended triangular lattices with large interaction ranges or on CG. In such cases, a direct test
of each potential bond for occupation becomes prohibitively expensive.

To overcome this difficulty, we adopt a fast sampling algorithm~\cite{Luijten1995} that directly
generates the sequence of occupied bonds, thereby avoiding explicit iteration over unoccupied ones.
Specifically, when traversing a sequence of bonds with occupation probability $p$, the probability that
the next occupied bond appears exactly $n$ steps after the current one follows a geometric distribution,
\begin{equation}
P(n) = (1-p)^{n-1} p,
\end{equation}
with cumulative distribution function $F(n) = 1 - (1-p)^n$. Given a uniform random number $u \in [0,1)$, the
interval $n$ to the next occupied bond is generated via the inverse transform
\begin{equation}
n = 1 + \left\lfloor \frac{\ln(1-u)}{\ln(1-p)} \right\rfloor,  
\end{equation}
where $\lfloor x\rfloor$ denotes the largest integer not exceeding $x$. This procedure allows us to skip all
intermediate unoccupied bonds and jump directly to the next occupied bond, and subsequently to further occupied
ones. As a result, the computational complexity is reduced from $O(E)$ to $O(p_c E)$, where $E$ is the
total number of potential bonds, leading to a substantial acceleration of the simulations when $p_c \ll 1$.

\subsection{Observables}   \label{sec-obs}

For each realization, we measure the following quantities: 
\begin{itemize}
\item Let $\mathcal{C}_1$ denote the size of the largest $k$-core cluster in a single realization. The mean
largest cluster size is defined as
\begin{equation}
C_1 \equiv \langle \mathcal{C}_1 \rangle,
\end{equation}
and the order parameter is given by
\begin{equation}
m \equiv \frac{C_1}{V}.
\end{equation}

\item The susceptibility-like quantity, characterizing the fluctuations of the order parameter, is defined as
\begin{equation}
\chi \equiv \frac{\langle \mathcal{C}_1^2 \rangle - \langle \mathcal{C}_1 \rangle^2}{V}.
\end{equation}

\item A critical polynomial defined as
\begin{equation}
P_B = R_2 - R_0,
\end{equation}
where $R_2$ and $R_0$ denote the wrapping probabilities of having a cluster wrapping along both directions
and having no wrapping cluster, respectively~\cite{Mertens2016,Xu2021}.
Across the transition, $R_2$ typically increases from $0$ to $1$, while $R_0$ decreases from $1$ to
$0$, so that $P_B \in [-1,1]$.

For systems belonging to the 2D percolation universality class, the wrapping probabilities satisfy
a universal topological relation at criticality, yielding $R_2 = R_0$. As a consequence, the
condition $P_B = 0$ provides a precise estimator of the critical point with significantly
reduced finite-size corrections.
\end{itemize}

Unless otherwise specified, $\langle \cdot \rangle$ and the wrapping
probabilities above are measured over all samples. For CGs and the
discontinuous regime, we also apply a conditional ensemble only on active samples, i.e., samples with $\mathcal{C}_1>0$.
To distinguish this ensemble from the conventional one, a subscript $+$
is used for the corresponding conditional observables:
\begin{align}
C_{1,+} & \equiv \langle \mathcal{C}_1\rangle_+,\\
m_+ & \equiv \frac{C_{1,+}}{V},          \\
\chi_+ &\equiv \frac{\langle \mathcal{C}_1^2\rangle_+ - \langle \mathcal{C}_1\rangle_+^2}{V},  \\
P_{B,+} &\equiv R_{2,+}-R_{0,+}.   
\end{align}
If no active samples are present, the $k$-core is taken to be empty, and each observable is assigned 
its value for $\mathcal{C}_1=0$.
This convention keeps the conditional average well defined on both sides of criticality.

\section{Hybrid $k$-core transition on complete graphs} \label{sec-cg}

\begin{figure}
\centering
\includegraphics[width=\linewidth]{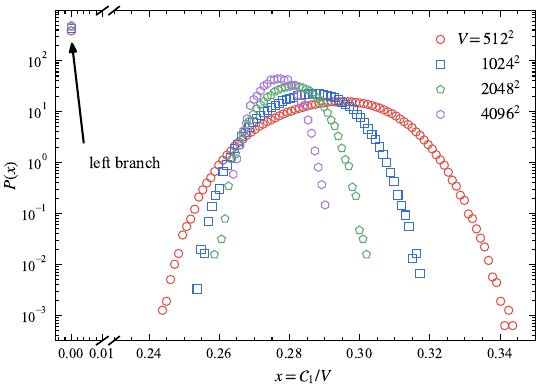}
\caption{Bimodal distribution $P(x)$ of the rescaled largest cluster size $x \equiv \mathcal{C}_1/V$ for $3$-core
percolation on CGs. All data are sampled at criticality, i.e., ER graphs with average
degree $q_c \approx 3.3509$. With increasing $V$, the active branch narrows rapidly
and its position approaches the theoretical jump value from above, reflecting the fast
finite-size convergence of the surviving-core samples on the CG.}
\label{f-1}
\end{figure}

Before turning to 2D lattices, we first consider $k$-core percolation on CGs, where the transition
is known to be hybrid for $k \ge 3$.
Specifically, the conditional order parameter $m_+$ exhibits a discontinuous
jump from $0$ to a finite value $m_0$, followed by a critical singularity when approaching the transition
from the supercritical side ($q \to q_c^+$),
\begin{align}
m_+ - m_0 &\sim (q - q_c)^{\beta}, \\
\chi_+ &\sim (q - q_c)^{-\gamma},
\end{align}
with mean-field critical exponents $\beta = 1/2$, $\gamma = 1$, and
$\nu = 2$~\cite{Chalupa1979,Pittel1996,Dorogovtsev2006,Goltsev2006}.

However, the finite-size behavior of such hybrid transitions is highly nontrivial.
The coexistence of a discontinuous jump and critical fluctuations leads to strong sample-to-sample
variability, making it difficult to simultaneously resolve both features in finite systems.
In general, for finite systems, different realizations at criticality can evolve into two qualitatively
distinct outcomes after the pruning process: an \emph{active} state, in which a macroscopic $k$-core
forms ($\mathcal{C}_1/V \approx m_0$), and an \emph{inactive} state,
in which the entire network is pruned ($\mathcal{C}_1=0$).
This coexistence gives rise to a bimodal distribution of the order parameter.

In Fig.~\ref{f-1}, we show the distribution $P(x)$ with $x = \mathcal{C}_1/V$ for $k=3$, for which the
critical average degree is $q_c \approx 3.3509$ and the jump amplitude is
$m_0 \approx 0.2676$~\cite{Chalupa1979,Pittel1996,Dorogovtsev2006,Goltsev2006}. Two distinct
branches are clearly observed. The left branch is essentially a $\delta$-function-like contribution
at $x=0$, corresponding to realizations without a $k$-core, while the right branch represents samples
with a nontrivial $k$-core. As $V$ grows, the width of the right branch decreases
quickly, so the fluctuations around the jump value become sharply localized.

\begin{figure}
\centering
\includegraphics[width=\linewidth]{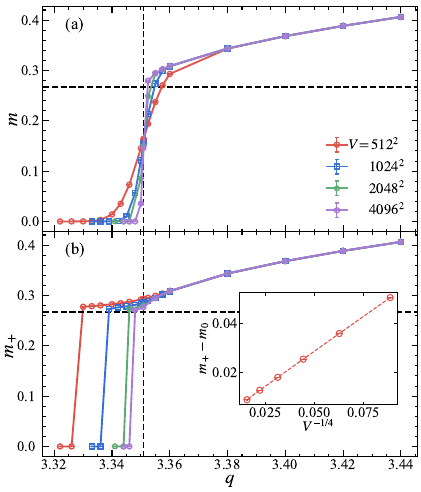}
\caption{Order parameter as a function of the average degree $q$ for $3$-core percolation on CGs.
Panel (a) shows the results averaged over all samples, which leads to a smeared, seemingly continuous
transition. Panel (b) shows the conditional average $m_+$ obtained using samples with $\mathcal{C}_1>0$, revealing a 
clear discontinuous
jump followed by a critical singularity. The slope near $q_c$ increases with system size,
in excellent agreement with theoretical predictions $m_0 \approx 0.2676$ and $q_c \approx 3.3509$ (dashed lines).
The inset of (b) plots $m_+(q_c,V)-m_0$ against $V^{-1/4}$, testing the finite-size form expected from $\beta/\nu=1/4$.}
\label{f-2}
\end{figure}

This bimodal distribution exists at any fixed average degree near $q_c$, and the weight of the finite-$x$ peak
increases gradually as $q$ increases. As a consequence, the ensemble average will smear out the intrinsic
discontinuous jump and produce an apparently continuous transition near $q_c$, as illustrated in
Fig.~\ref{f-2} (a). The all-sample curves show an approximate crossing whose
position tends toward $(q_c,m_0/2)$ as the system size increases. This crossing reflects
the finite-size bimodal distribution. Near the transition, collapsed samples contribute
$m=0$, while samples with a surviving macroscopic $k$-core contribute a value close to
$m_0$, and the two statistical weights become comparable. The critical part of the hybrid
transition is instead seen in the conditional ensemble, where the post-jump singularity
and the divergent fluctuations are retained.

To recover the intrinsic behavior of the hybrid transition, we adopt the conditional averaging procedure:
realizations in which the giant $k$-core fails to form ($\mathcal{C}_1=0$) are excluded, and statistical averages are
taken only over the active ensemble. In practice, this amounts to restricting the averages to samples
with $\mathcal{C}_1 > 0$. We denote the resulting order parameter by $m_+$.

As shown in Fig.~\ref{f-2} (b), this procedure restores a sharp discontinuous jump of the order parameter at
the transition. In the subcritical regime, $m_+$ remains zero, whereas approaching criticality from the supercritical
side, the numerical data converge to a finite plateau value $m_0$, consistent with the theoretical prediction
(horizontal dashed line). Meanwhile, with increasing system size, the location of the sharp increase in $m_+$
systematically approaches the theoretical critical value (vertical dashed line).
At the critical point $q_c$, the finite-size correction follows
$m_+(q_c,V)-m_0\sim V^{-\beta/\nu}=V^{-1/4}$, as shown in the inset of Fig.~\ref{f-2}(b).
This provides direct evidence that the system exhibits critical behavior
when approaching the transition point from the supercritical side,
thereby confirming the hybrid nature of the transition.
At the same time, it validates the effectiveness of the conditional averaging method
in correctly extracting the critical behavior of the hybrid transition.

\begin{figure}
\centering
\includegraphics[width=\linewidth]{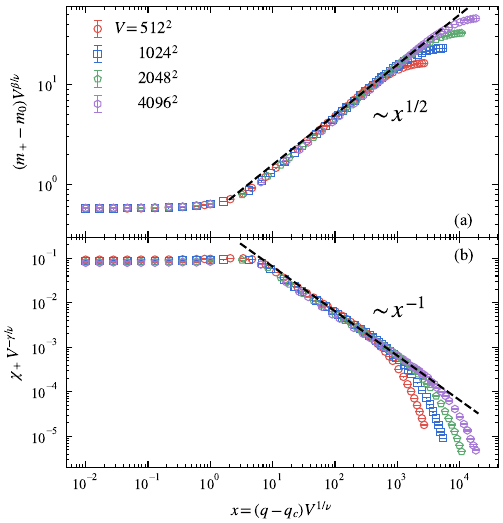}
\caption{Finite-size behaviors of the conditional order parameter $m_+$ and susceptibility $\chi_+$ near criticality of
$3$-core
percolation on CGs. The data are obtained by restricting the averages to samples with $\mathcal{C}_1 > 0$. Excellent
data collapse is achieved using the mean-field exponents $\beta=1/2$, $\gamma=1$, and $\nu=2$. The lines indicate
the scalings, $\tilde{m}(x) \sim x^{\beta}, \tilde{\chi}(x) \sim x^{-\gamma}$, for $x\gg1$.}
\label{f-3}
\end{figure}

Having established the appropriate statistical treatment, we next examine whether the critical singularities on the
supercritical side are correctly reproduced. The finite-size scaling forms of the order parameter and susceptibility
are expected to obey
\begin{align}
m_+(q,V) - m_0 &= V^{-\beta/\nu}\,\tilde{m}(x), \\
\chi_+(q,V) &= V^{\gamma/\nu}\,\tilde{\chi}(x),
\end{align}
where $x \equiv (q - q_c)V^{1/\nu}$ is the rescaled distance-to-criticality. As shown in Fig.~\ref{f-3},
plotting $(m_+-m_0)V^{\beta/\nu} = \tilde{m}(x)$ and $\chi_+ V^{-\gamma/\nu} = \tilde{\chi}(x)$ as functions
of $x$ in the supercritical regime yields
an excellent data collapse for different system sizes using the mean-field exponents $\beta=1/2$, $\gamma=1$,
and $\nu=2$. This reinforces that the critical singularities above $q_c$ are accurately captured by our simulations.

In the crossover critical regime with $x \gg 1$, the scaling functions exhibit the expected asymptotic power-law 
behaviors (see Fig.~\ref{f-3}),
\begin{align}
\tilde{m}(x) &\sim x^{\beta},  \\
\tilde{\chi}(x) &\sim x^{-\gamma}.
\end{align}
These results are consistent with the crossover finite-size scaling behavior of continuous phase
transitions~\cite{Li2024}. Taken together, this analysis provides a stringent validation of our numerical
procedures and establishes a reliable reference for comparison with the 2D lattice
results presented below.

\section{Critical behaviors of $k$-core percolation in two dimensions} \label{sec-2d}

\begin{figure}
\centering
\includegraphics[width=\linewidth]{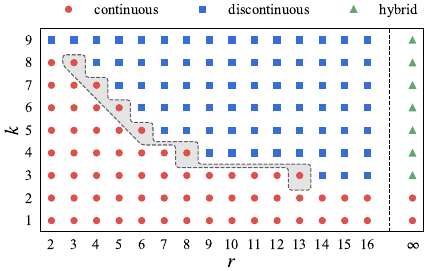}
\caption{Phase diagram of $k$-core percolation on the extended triangular lattice in the $(k,r)$ parameter space.
A special line $(k_s,r_s)$ separates a continuous transition regime (red circles), which belongs to the universality
class of standard 2D percolation, from a first-order regime (blue squares). No evidence of a hybrid transition
is found in two dimensions. The phase diagram is obtained from simulations with system size up to $L=4096$. Points in
the
shaded region, while suggestive of a continuous transition at the current level of precision, cannot be
unambiguously classified due to strong finite-size effects. For $r\to\infty$, the system reduces
to the CG, for which the hybrid transition can be observed for $k\geq3$ (green triangles).}
\label{f-4}
\end{figure}

We now turn to the 2D case, namely triangular lattices with a tunable but finite interaction range $r$.
Before presenting detailed numerical results, we first summarize our main findings through the phase
diagram in the $(k,r)$ plane (Fig.~\ref{f-4}). In stark contrast to the hybrid transition
observed in infinite-dimensional systems, the 2D system exhibits a special line $(k_s,r_s)$ that
separates two qualitatively distinct regimes. Below $(k_s,r_s)$, the transition is continuous
and belongs to the universality class of standard 2D percolation. Above $(k_s,r_s)$, the transition
becomes first-order, with no evidence of a genuine hybrid transition.

Due to strong finite-size effects in the crossover region and limited numerical resolution, some parameter
points (gray shaded region in Fig.~\ref{f-4}) cannot be unambiguously classified. Nevertheless,
these ambiguous cases do not affect our central conclusion, as they can only correspond to either
continuous or first-order transitions. Only in the limit $r\to\infty$ does the system reduce to the
CG, where the hybrid transition can be observed.

\subsection{Continuous transition regime}

\subsubsection{$r=1$}

We begin with $k$-core percolation on the standard triangular lattice ($r=1$). For $k=1$, the model
reduces to standard bond percolation, whose percolation threshold is exactly known as
$p_c = 2\sin(\pi/18) \approx 0.347296$~\cite{Sykes1964}, with correlation-length exponent
$\nu = 4/3$ and fractal dimension $d_f = 91/48$~\cite{Nienhuis1984}.

\begin{figure}
\centering
\includegraphics[width=\linewidth]{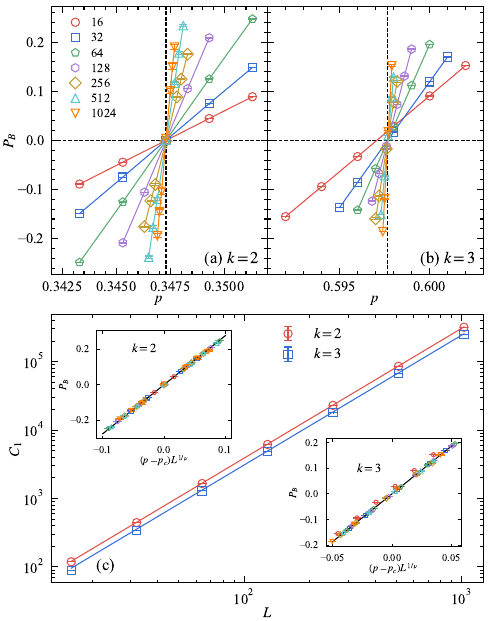}
\caption{Finite-size scaling of $k$-core percolation on triangular lattices ($r=1$). Panels
(a) and (b) show the critical polynomial $P_B$ as a function of the occupation probability $p$
near the transition for $k=2$ and $k=3$, respectively. The curves for different system sizes $L$
intersect at well-defined points, allowing for a precise determination of the thresholds $p_c$.
Panel (c) shows the size of the largest $k$-core cluster $C_1$ at $p_c$ as a function of $L$, where
the slope yields a fractal dimension $d_f \approx 1.8958$, consistent with the exact value $d_f=91/48$
for standard 2D percolation. The upper and lower insets in (c) display the data collapse of $P_B$
for $k=2$ and $k=3$, respectively, using the scaling ansatz in Eq.~(\ref{eq-pb}) and the exponent
$\nu = 4/3$.}  \label{f-5}
\end{figure}

We focus on the nontrivial cases $k=2$ and $k=3$. To locate the critical point, we analyze the
finite-size behavior of the critical polynomial $P_B = R_2 - R_0$. As shown in Fig.~\ref{f-5}(a,b),
the $P_B$ curves for different system sizes exhibit clean crossings at $P_B=0$, enabling a precise
determination of the percolation threshold. Quantitatively, we fit the data using the finite-size
scaling form
\begin{equation}
P_{B}(p, L) = P_{B,c} + a_{1}(p - p_c) L^{1/\nu}.
\label{eq-pb}
\end{equation}
For all finite-size scaling fits performed in this work, we apply a lower system-size cutoff
$L \ge L_{\rm min}$ to systematically suppress higher-order finite-size corrections. We assess
the goodness of fit by examining the reduced chi-square $\chi^2$ per degrees of freedom (DF). The
optimal $L_{\rm min}$ is chosen as the smallest value beyond which $\chi^2/{\rm DF}$ no longer
decreases significantly (i.e., by more than one unit) with further decreases in $L_{\rm min}$. A fit
is considered acceptable if $\chi^2/{\rm DF} \approx 1$. Furthermore, systematic uncertainties are carefully
estimated by comparing results from fits with varying correction terms.

Applying this procedure to Eq.~(\ref{eq-pb}) with $P_{B,c}=0$
yields $p_c = 0.347298(2)$ and $1/\nu = 0.749(2)$ for $k=2$, and $p_c = 0.5976774(7)$ and
$1/\nu = 0.750(4)$ for $k=3$. These results show that the $2$-core percolates at essentially the same
threshold as standard percolation, whereas the emergence of a giant $3$-core requires a higher bond
occupation probability.

Despite this difference in thresholds, both cases exhibit the same correlation-length exponent $\nu = 4/3$ as
standard 2D percolation. As shown in the insets of Fig.~\ref{f-5}(c), the data for $P_B$ at different system
sizes collapse well when plotted against $(p-p_c)L^{1/\nu}$ using $\nu=4/3$, providing strong evidence that both
transitions belong to the same universality class.

For standard 2D percolation, the largest cluster at criticality scales as $C_1 \sim L^{d_f}$ with $d_f=91/48$.
In Fig.~\ref{f-5} (c), we show that the largest $k$-core at criticality follows the same scaling behavior.
Fitting the data using the finite-size scaling ansatz
\begin{equation}
C_1 = L^{d_f} (a_0 + a_1 L^{-\omega_1}),
\label{eq-c1}
\end{equation}
we obtain $d_f = 1.8958(3)$ for $k=2$ and $d_f = 1.8958(6)$ for $k=3$, both in excellent agreement with the exact
value $d_f = 91/48 \approx 1.8958$.

Taken together, these results demonstrate that the $k$-core percolation transitions for $k=2$ and $k=3$ on the
triangular lattice are continuous and belong to the universality class of standard 2D percolation. For $k=2$,
this is consistent with earlier field-theoretic analyses~\cite{Harris1983,Harris1983a}. More importantly, our
results show that even under the stronger local constraint imposed by $k=3$, which requires cooperative local
structures to sustain the surviving core, the universality class remains unchanged.

Finally, due to the limited coordination number $z=6$, nontrivial $k$-core clusters cannot survive for $k>3$
once a finite fraction of bonds is removed. To systematically explore larger values of $k$ while retaining
nontrivial transitions, we therefore turn to triangular lattices with extended interaction ranges $r>1$.

\begin{table}
\centering
\caption{Fitted critical points $q_c$ and correlation-length exponents $\nu$ for various $(k,r)$ pairs
in the continuous transition regime. The fits use the ansatz in Eq.~(\ref{eq-pb}) with $P_{B,c}=0$.
The exponents are consistent with the 2D percolation value $\nu = 4/3$. Entries without errors
represent fixed values in the corresponding fit.}
\label{tab3}
\begin{ruledtabular}
\begin{tabular}{clllll}
\multicolumn{1}{c}{$(k,r)$}   &  \multicolumn{1}{c}{$L_{\rm min}$} &  \multicolumn{1}{c}{$q_c$} & \multicolumn{1}{c}{$1/\nu$} & \multicolumn{1}{c}{$a_1$}  & \multicolumn{1}{c}{$\chi^2/{\rm DF}$}  \\
    \hline
  (4,4) &   256   &  5.308352(2)  &  0.742(2)    &  0.554(6)      &  45.23/32   \\
        &   512   &  5.308352(2)  &  0.749(3)    &  0.53(1)      &  35.66/27   \\
        &   512   &  5.308352(2)  &  0.75    &  0.523(1)      &  35.87/28   \\
   \hline
  (4,5)
   &   512  &  5.340505(2)  &  0.739(4)  &  -0.63(2)  &  39.18/27   \\
  &   1024  &  5.340504(2)  &  0.754(6)  &  -0.57(2)  &  23.89/22   \\
  &   1024  &  5.340504(2)  &  0.75  &  -0.582(1)  &  24.31/23   \\
   \hline
  (4,6)
 &   1536  &  5.368325(2)  &  0.74(1)  &  -0.67(7)  &  26.12/17   \\
  &   2048  &  5.368324(2)  &  0.75(2)  &  -0.66(9)  &  12.91/12   \\
  &   1536  &  5.368325(2)  &  0.75  &  -0.639(2)  &  26.52/18   \\
   \hline
  (5,3)
  &    256  &  7.062337(2)  &  0.742(1)  &  -0.494(4)  &  31.59/32   \\
  &   512  &  7.062336(2)  &  0.747(3)  &  -0.47(1)  &  26.51/27   \\
  &   512  &  7.062336(2)  &  0.75  &  -0.4663(9)  &  27.08/28   \\
   \hline
  (5,4)
  &     512  &  7.133873(2)  &  0.742(3)  &  -0.54(1)  &  23.85/27   \\
  &   1024  &  7.133874(2)  &  0.752(5)  &  -0.50(2)  &  14.64/22   \\
  &   1024  &  7.133874(1)  &  0.75  &  -0.508(1)  &  14.77/23   \\
   \hline
  (5,5)
  &  1024  &  7.197967(3)  &  0.726(7)  &  -0.66(4)  &  35.00/22   \\
  &   1536  &  7.197971(2)  &  0.73(1)  &  -0.64(5)  &  17.08/17   \\
  &   1536  &  7.197971(2)  &  0.75  &  -0.545(2)  &  20.83/18   \\
   \hline
  (6,2)
 &   256  &  8.685196(3)  &  0.737(3)  &  -0.539(0)  &  33.83/33   \\
  &   512  &  8.685195(3)  &  0.744(7)  &  -0.51(2)  &  28.27/27   \\
  &   512  &  8.685195(3)  &  0.75  &  -0.489(2)  &  29.10/28   \\
   \hline
  (6,3)
 &  512  &  8.809838(3)  &  0.740(6)  &  -0.49(2)  &  29.45/28   \\
 &    1024  &  8.809838(3)  &  0.758(7)  &  -0.43(2)  &  15.24/22   \\
  &   1024  &  8.809838(3)  &  0.75  &  -0.452(2)  &  16.10/23   \\
   \hline
  (6,4)
  &    1536  &  8.931353(3)  &  0.74(2)  &  -0.49(6)  &  19.93/17   \\
  &   2048  &  8.931352(4)  &  0.75(2)  &  -0.48(9)  &  16.12/12   \\
  &   1536  &  8.931353(3)  &  0.75  &  -0.465(2)  &  20.15/18   \\
\end{tabular}
\end{ruledtabular}
\end{table}

\begin{table}
\centering
\caption{Fitted fractal dimensions $d_f$ for various $(k,r)$ pairs. The fits use Eq.~(\ref{eq-c1}). All values
are in good agreement with the standard 2D percolation exponent $d_f \approx 1.896$. Entries without errors
represent fixed values in the corresponding fit.}
\label{tab4}
\begin{ruledtabular}
\begin{tabular}{cllllll}
\multicolumn{1}{c}{$(k,r)$}   &  \multicolumn{1}{c}{$L_{\rm min}$} &  \multicolumn{1}{c}{$d_f$} & \multicolumn{1}{c}{$a_0$} & \multicolumn{1}{c}{$\omega_1$} & \multicolumn{1}{c}{$a_1$} & \multicolumn{1}{c}{$\chi^2/{\rm DF}$}  \\
    \hline
  (4,4)
        & 256    &  1.896(9)   &  0.450(4)  &  1.5(2)  &  -35(9)  &  0.58/3   \\
        & 256    &  1.8961(3)  &  0.4494(9)  &  1.5  &  -38(1)  &  0.59/4   \\
        & 512    &  1.8960(5)  &  0.450(2)  &  1.5  &  -40(5)  &  0.57/3   \\
   \hline
  (4,5)
   & 256  &    1.891(6)  &  0.49(3)  &  1.0(3)  &  -7(9)  &  5.13/3   \\
   & 512  &    1.895(1)  &  0.471(5)  &  1.5  &  -122(5)  &  2.88/3   \\
   & 512  &    1.891(2)  &  0.490(9)  &  1.0  &  -9(1)  &  4.69/3   \\
   & 1024  &    1.898(4)  &  0.46(1)  &  1.0  &  -2(3)  &  1.42/2   \\
   \hline
  (4,6)
   & 512  &    1.894(2)  &  0.519(9)  &  1.0  &  -16(1)  &  1.90/3   \\
   & 512  &    1.902(1)  &  0.484(6)  &  1.5  &  -229(9)  &  1.74/3   \\
   & 1024  &    1.897(5)  &  0.51(2)  &  1.0  &  -13(5)  &  1.62/2   \\
   & 1024  &    1.901(4)  &  0.49(2)  &  1.5  &  -243(90)  &  1.72/2   \\
   \hline
  (5,3)
   & 256  &    1.894(1)  &  0.549(5)  &  1.2(2)  &  -10(8)  &  0.35/3   \\
   & 256  &    1.8923(4)  &  0.557(2)  &  1.0  &  -4.5(1)  &  0.52/4   \\
   & 256  &    1.8955(3)  &  0.541(1)  &  1.5  &  -49(2)  &  0.77/4   \\
   & 512  &    1.8948(4)  &  0.545(2)  &  1.5  &  -60(5)  &  0.28/3   \\
   \hline
  (5,4)
   & 512  &    1.891(2)  &  0.60(1)  &  1.0  &  -13(2)  &  3.39/3   \\
   & 512  &    1.896(2)  &  0.575(7)  &  1.5  &  -186(21)  &  3.12/3   \\
   & 1024  &    1.893(6)  &  0.59(3)  &  1.0  &  -10(6)  &  2.98/2   \\
   & 1024  &    1.897(5)  &  0.57(2)  &  1.5  &  -179(117)  &  3.12/2   \\
   \hline
  (5,5)
   & 1024  &    1.889(2)  &  0.75(2)  &  0.5  &  -2.3(2)  &  0.13/2   \\
   & 1024  &    1.913(1)  &  0.587(6)  &  1.5  &  -407(4)  &  0.19/2   \\
   & 1024  &    1.907(1)  &  0.622(7)  &  1.0  &  -23(2)  &  0.13/2   \\
    \hline
  (6,2)
   &  256  &    1.8901(6)  &  0.671(3)  &  1.0  &  -8.6(3)  &  1.04/4   \\
   &  256  &    1.8951(6)  &  0.642(3)  &  1.5  &  -92(4)  &  2.07/4   \\
   &  512  &    1.890(1)   &  0.669(6)  &  1.0  &  -8.3(9)  &  1.00/3   \\
   &  512  &    1.8936(6)  &  0.650(3)  &  1.5  &  -118(9)  &  0.48/3   \\
   &  1024  &    1.894(2)  &  0.647(9)  &  1.0  &  -4(2)  &  0.26/2   \\
   &  1024  &    1.895(1)  &  0.641(6)  &  1.5  &  -70(34)  &  0.24/2   \\
   \hline
   (6,3)
    &  256  &    1.895(2)  &  0.65(1)  &  1.0  &  -9.8(9)  &  8.45/4   \\
    &  512  &    1.892(3)  &  0.67(2)  &  1.0  &  -13(3)  &  5.50/3   \\
    &  512  &    1.896(2)  &  0.644(9)  &  1.5  &  -184(28)  &  3.54/3   \\
    &  1024  &    1.903(3)  &  0.61(2)  &  1.0  &  1(3)  &  0.55/2   \\
    \hline
    (6,4)
    &  1024  &    1.91(1)  &  0.67(6)  &  1.0  &  -15(4)  &  6.68/2   \\
    &  1024  &    1.90(2)  &  0.8(1)  &  0.5  &  -2(1)  &  6.23/2   \\
\end{tabular}
\end{ruledtabular}
\end{table}

\subsubsection{$r>1$}

\begin{figure*}
\centering
\includegraphics[width=\linewidth]{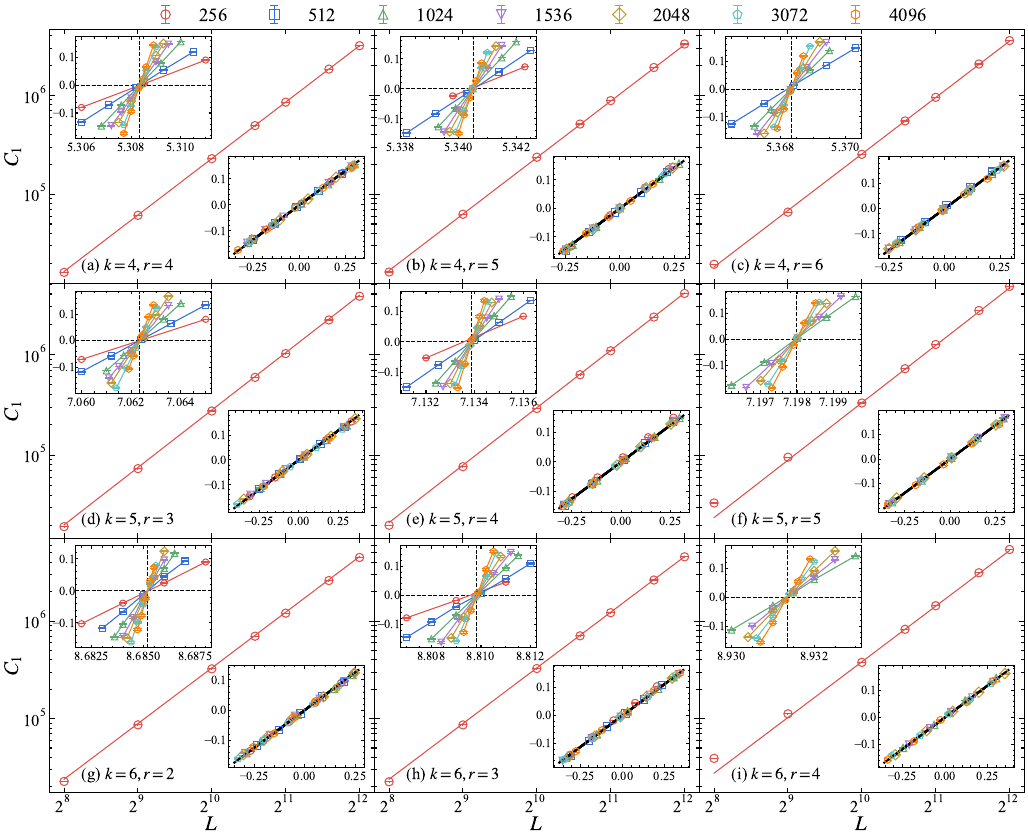}
\caption{Universality of the continuous phase transitions for representative $(k,r)$ pairs with
$k=4, 5, 6$. The panels (a)-(i) display the power-law scaling of the largest-cluster size $C_1$
at the critical point $q_c$ (see Table~\ref{tab3}). The solid red lines represent the scaling
$C_1\sim L^{d_f}$ with $d_f = 91/48$. The upper-left insets in each panel depict the critical
polynomial $P_B$, which exhibits a clear crossing point at $P_B=0$. The lower-right insets show
the data collapse of $P_B$ using the correlation-length exponent $\nu = 4/3$, further confirming
that the transition belongs to the 2D percolation universality class. Notably, finite-size effects
become increasingly pronounced as $k$ and $r$ increase. To account for this, the minimum system size
used in the analysis was progressively increased, reaching $L=512$ for $k=4$ and $L=1024$ for $k=5, 6$
in the rightmost column.}     \label{f-6}
\end{figure*}

Upon increasing the interaction range $r$, the coordination number $z(r)$ grows, enabling nontrivial
$k$-core percolation transitions at finite thresholds even for larger values of $k$. In the
continuous-transition regime ($k < k_s$), the critical point can again be accurately identified
from the crossing of the critical polynomial $P_B$, similar to Fig.~\ref{f-5} (a,b).

For large $r$, the critical bond occupation probability $p_c$ becomes very small. In this regime,
it is more convenient to use the average degree $q=pz(r)$ as the control parameter, so that
the transition is characterized by a critical value $q_c=p_c z(r)$, as in CGs.

Representative results for different $(k,r)$ pairs within the continuous regime are shown in
Fig.~\ref{f-6}. The upper insets of each panel display clear crossings of $P_B$, indicating
well-defined continuous transitions. By fitting the data to the finite-size scaling form in
Eq.~(\ref{eq-pb}), we extract the critical points $q_c$ and the correlation-length exponent $\nu$,
summarized in Table~\ref{tab3}. For all cases, we obtain $\nu = 4/3$ within numerical accuracy,
consistent with the universality class of standard 2D percolation. This conclusion is further
supported by data collapse analyses: plotting $P_B$ as a function of $(q - q_c)L^{1/\nu}$ with
$\nu = 4/3$ yields excellent collapse across all system sizes, as shown in the lower
insets of Fig.~\ref{f-6}.

In addition, the size of the largest $k$-core at criticality follows the scaling
form $C_1 \sim L^{d_f}$. Fitting the data using Eq.~(\ref{eq-c1}) yields fractal
dimensions $d_f$ listed in Table~\ref{tab4}, all of which agree with the exact 2D
percolation value $d_f = 91/48$. Taken together, these results demonstrate that
throughout the continuous-transition regime, $k$-core percolation on triangular
lattices with finite interaction range remains in the universality class of
standard 2D percolation, independent of $r$.

As the interaction range $r$ increases, the threshold $k_s$ gradually approaches the
interval $2 < k_s < 3$. Field-theoretic analyses predict that the $k=2$ case always
belongs to the standard percolation universality class~\cite{Harris1983,Harris1983a},
while renormalization-group arguments indicate that finite interaction ranges do not alter
the critical fixed point. Consequently, $k_s$ is not expected to decrease below $2$ for
any finite $r$, in agreement with our numerical observations.

\begin{figure}
\centering
\includegraphics[width=\linewidth]{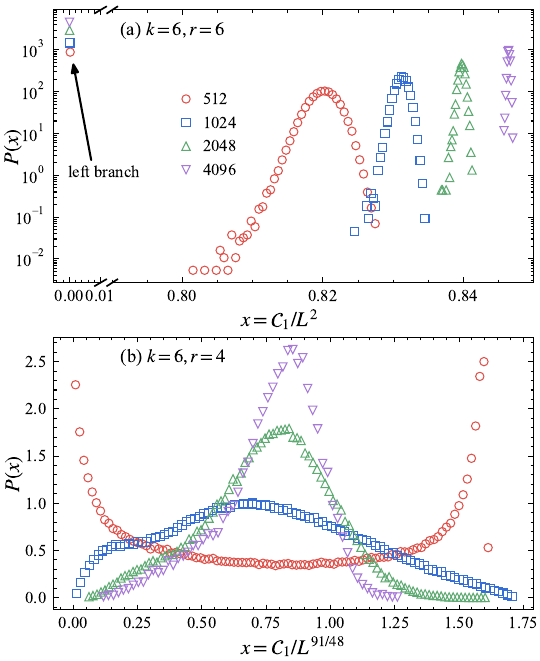}
\caption{Probability distribution $P(x)$ of the rescaled largest cluster size at
the critical point. (a) Discontinuous phase transition regime ($k=6,r=6$). $P(x)$
exhibits a distinct bimodal character with one peak at zero and another at a finite
value. (b) Continuous phase transition regime ($k=6,r=4$). As the system size $L$
increases, the distribution $P(x=C_1/L^{d_f})$ evolves from a double-peak structure
to a single-peak form.} \label{f-7}
\end{figure}

\begin{figure}
\centering
\includegraphics[width=\linewidth]{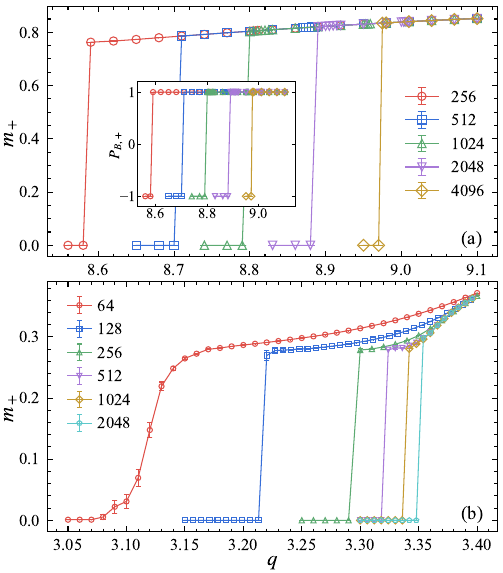}
\caption{Discontinuous transition analysis for large $k$ and $r$ calculated using
conditional averaging over samples with $\mathcal{C}_1>0$ for different system sizes $L$. Panel (a)
shows the conditional order parameter $m_+$ as a function of the average degree $q$ for ($k=6,r=6$).
The curve exhibits a sharp, discontinuous jump from zero to a finite value at the
transition point. Crucially, immediately following the jump, the order parameter
increases smoothly without exhibiting the power-law singularity characteristic
of hybrid transitions found in network models. The inset of (a) displays the
critical polynomial $P_B$, which transitions abruptly from $-1$ to $1$,
behaving like a step function. This lack of critical fluctuations confirms
the nature of the transition as a first-order transition. Panel (b)
shows the case $(k,r)=(3,64)$. For finite system sizes, the interaction
range is sufficiently large that the system exhibits an apparent crossover
toward CG behavior, and hybrid-like signatures can be observed. However, since $r$ remains
finite, these features are only transient finite-size effects: as $L$ increases, the
critical-like behavior gradually disappears, and the transition crosses over
to a first-order one.}      \label{f-8}
\end{figure}

\begin{figure}
\centering
\includegraphics[width=\linewidth]{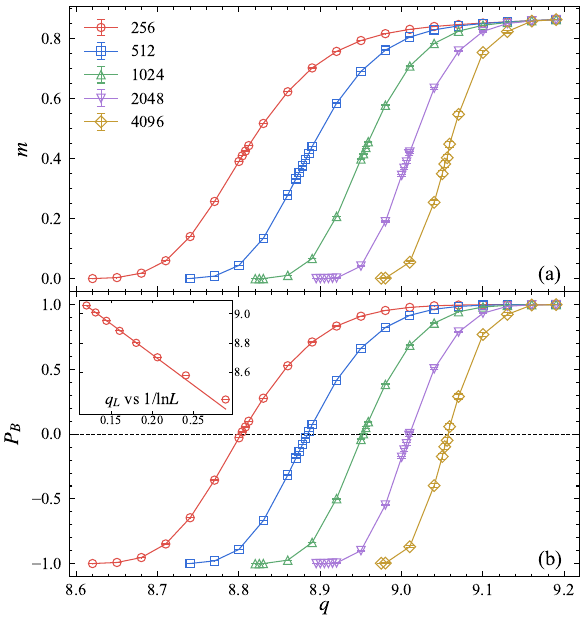}
\caption{Discontinuous transition analysis for the representative parameter $(k=6,r=6)$
by averaging over all samples for different system sizes $L$. Panels (a) and
(b) display the order parameter $m$ and the critical polynomial $P_B$ as
functions of the average degree $q$. Although the curves appear continuous,
$P_B$ does not exhibit a fixed crossing point; instead, the curves shift to
the right as the system size $L$ increases. The effective transition point
$q_L$, defined by $P_B(q_L)=0$, is plotted against $1/\ln L$ in the inset
of (b), revealing a linear scaling trend consistent with a
nucleation-driven mechanism.} \label{f-9}
\end{figure}

\subsection{First-order transition regime}

\subsubsection{Bimodal distribution and absence of critical fluctuations}

In the first-order transition regime (blue squares in Fig.~\ref{f-4}), the transition
is no longer governed by critical fluctuations. Instead, the system undergoes an abrupt switch
between two macroscopic states: an active phase with a finite $k$-core
($\mathcal{C}_1>0$), and an inactive
phase in which the entire structure is completely pruned ($\mathcal{C}_1=0$). A direct signature of this
behavior is the bimodal distribution of the order parameter at the transition point. As shown
in Fig.~\ref{f-7}(a) for the representative case $(k,r)=(6,6)$, the distribution exhibits
two well-separated peaks: a $\delta$-function-like contribution at $m=0$, corresponding to
the inactive phase, and a second peak at finite $m$, associated with the active phase.
With increasing system size, the finite-$m$ peak shifts toward larger values,
in clear contrast to the behavior on CGs (see Fig.~\ref{f-1}).
Moreover, although the finite-$m$ peak also becomes narrower with increasing $L$,
this narrowing is substantially slower than in the CG case.
These differences further indicate that the discontinuous transition here is fundamentally distinct
from the hybrid transition observed on CGs.

For comparison, Fig.~\ref{f-7}(b) shows a representative case in the continuous regime, $(k,r)=(6,4)$.
Although a bimodal structure appears for small system sizes, it gradually evolves into a unimodal distribution
as $L$ increases, consistent with a continuous transition subject to strong finite-size effects. In practice,
this qualitative evolution of $P(x)$ provides a useful indication for distinguishing continuous and discontinuous
transitions. For some parameter sets with large $r$, however, this evolution becomes extremely slow with
increasing $L$. Within the accessible system sizes, only the tendency toward a unimodal form can be
identified, while a fully merged single peak is not yet reached. These cases are therefore
indicated by the shaded region in Fig.~\ref{f-4}.

To isolate the intrinsic behavior of the ordered phase, we perform conditional averaging over
realizations with $\mathcal{C}_1>0$. As shown in Fig.~\ref{f-8}(a), this procedure restores a sharp
discontinuous jump of the conditional order parameter $m_+$. However, in contrast to hybrid transitions, the
post-jump behavior is smooth and does not exhibit any critical singularity as $q \to q_c^{+}$.
Consistently, under conditional averaging, $P_B$ behaves as an almost ideal step function,
jumping directly from $-1$ to $1$ [see inset of Fig.~\ref{f-8}(a)], with no signature of
critical fluctuations. This indicates that the transition is
first-order, rather than hybrid as in Fig.~\ref{f-2}(a).

For finite systems with sufficiently large $r$, the interaction range may exceed the
relevant correlation scale, so that the system can temporarily display CG-like
behavior~\cite{Xue2024}. In this pre-asymptotic regime, apparent hybrid-transition
signatures may therefore emerge as crossover effects between 2D and infinite-dimensional
behavior. However, for any finite $r$, these critical-like features disappear as $L$
increases, and the transition ultimately crosses over to a first-order one. A
representative example for $r=64$ is shown in Fig.~\ref{f-8}(b).

\subsubsection{Pseudocritical point}

If all realizations are averaged indiscriminately, the coexistence of the two phases leads to a strong smearing of 
the order parameter. As shown in Fig.~\ref{f-9}(a), the curve $m(q)$ appears continuous and drifts systematically 
with increasing system size. This behavior should be contrasted with the CG case shown in Fig.~\ref{f-2}(a), 
where the transition sharpens toward a well-defined critical point as the system size increases. Similar 
behavior is observed for the critical polynomial $P_B$. As shown in Fig.~\ref{f-9}(b), the curves for 
different system sizes do not exhibit a size-independent crossing; instead, the entire curve shifts 
systematically with $L$.

Based on these drifting curves, we define a pseudocritical point $q_L$ through the condition $P_B(q_L)=0$. 
We find that $q_L$ follows a logarithmic finite-size scaling form,
\begin{equation}
q_L = q_c + \frac{a}{\ln L},        \label{eq-qL}
\end{equation}
as shown in the inset of Fig.~\ref{f-9}(b). For the representative case $(k,r)=(6,6)$, fitting the data 
to Eq.~(\ref{eq-qL}) yields $q_c \simeq 9.55(2)$.

Notably, the drifting curves in Figs.~\ref{f-9} and \ref{f-8} remain far from the estimated critical 
point $q_c \simeq 9.55(2)$, where both $m$ and $P_B$ become nearly independent of $L$ and no critical 
behavior can be observed. This is fundamentally different from the hybrid transition on CGs, where the 
order parameter on the ordered side, $m_+(q_c,V)$, approaches the jump value with a power-law finite-size 
correction $V^{-1/4}$, as shown in the inset of Fig.~\ref{f-2}(b). The absence of such critical 
scaling further supports the interpretation that the transition in two dimensions is first-order rather than hybrid.

Moreover, the scaling form in Eq.~(\ref{eq-qL}) differs fundamentally from that of conventional thermodynamic 
first-order transitions, where the finite-size shift typically scales as $L^{-d}$. Instead, the logarithmic 
drift points to a nucleation-driven mechanism, analogous to bootstrap percolation~\cite{Holroyd2003}. In 
this picture, the system remains metastable until a rare critical void nucleates and triggers a cascade 
that destroys the $k$-core. By contrast, the hybrid transition on CGs is still governed by power-law 
critical scaling~\cite{Gao2024}. Taken together, the absence of critical singularities and the 
logarithmic finite-size drift provide compelling evidence that the transition in this regime is 
a genuine first-order transition rather than a hybrid one.

\section{Renormalization-flow interpretation} \label{sec-rfi}

\begin{figure}
\centering
\includegraphics[width=\linewidth]{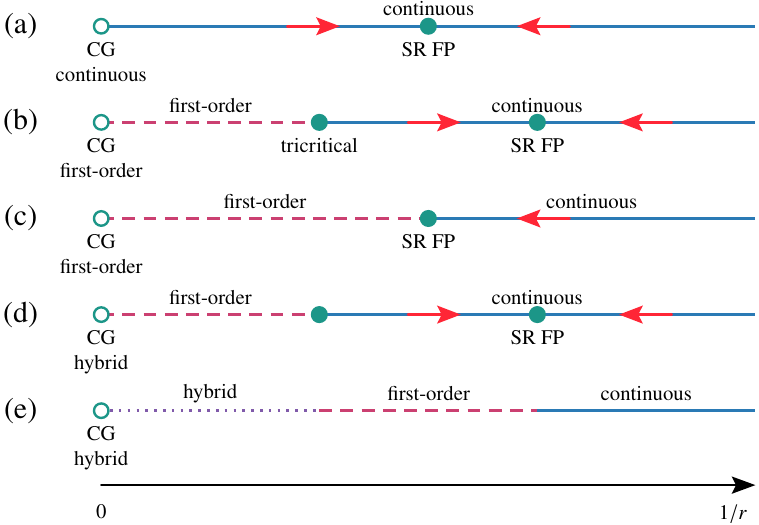}
\caption{Schematic renormalization-flow diagrams as the interaction range $r$ is varied. The horizontal coordinate 
is $1/r$, such that the CG mean-field limit corresponds to $1/r=0$. Panels (a)--(c) illustrate generic scenarios for 
connecting mean-field and finite-range 2D behaviors in the $Q$-state Potts model. (a) $Q=1$ and $2$: the 
transition remains continuous for any $r$. The mean-field fixed point is unstable against finite-range 
perturbations, while the short-range (SR) 2D fixed point (FP) is stable. This diagram is also applied to $k$-core 
percolation with $k=1$ and $2$. (b) $Q=3$: an unstable tricritical point appears at finite $r$, 
separating the short-range 2D regime from the first-order regime connected to the CG limit. (c) $Q=4$: 
similar to (b), except that the tricritical and short-range 2D fixed points merge. (d) 
Renormalization-flow scenario suggested by the present work for $k$-core percolation with 
$k\ge3$. The hybrid transition at the CG limit is unstable under finite-range perturbations 
and becomes strictly first-order for finite but sufficiently large interaction ranges. For 
sufficiently short interaction ranges, the system instead belongs to the standard 2D 
percolation universality class. (e) Alternative scenario proposed in Ref.~\cite{Xue2024}, 
where increasing $r$ drives the system successively through continuous, first-order, and 
hybrid regimes before reaching the CG limit.}
\label{f-10}
\end{figure}

More generally, the crossover from infinite to finite interaction range is known to be highly nontrivial in statistical
systems. A unified understanding of the present results can be obtained from the perspective of renormalization flows
(Fig.~\ref{f-10}).

A prototypical example is the 2D $Q$-state Potts model with an extended interaction range $r$. For $Q=1$, the model
reduces to standard percolation, corresponding to $k$-core percolation with $k=1$. As illustrated schematically in
Fig.~\ref{f-10}(a), the limit $1/r \to 0$ corresponds to the CG limit, which belongs to the mean-field percolation
universality class. However, this mean-field fixed point is unstable with respect to finite-range
perturbations~\cite{Ouyang2018}: all systems with finite $r$ ultimately flow to the short-range 2D percolation fixed
point. This renormalization-flow structure is physically natural, since any system with finite $r$ remains 2D, and
therefore any continuous transition should belong to the corresponding 2D universality class. For $Q=2$, corresponding
to the Ising model, the renormalization flow has the same qualitative structure~\cite{Luijten1996,Luijten1997}.

The situation changes for $Q=3$, as illustrated in Fig.~\ref{f-10}(b). In the CG limit, the transition becomes
first-order, rendering the mean-field fixed point stable. Consequently, systems with sufficiently large but finite $r$
can also exhibit first-order transitions~\cite{Qian2016}. At the same time, for small $r$, the transition remains
continuous and belongs to the short-range 2D universality class. These two regimes are separated by a tricritical point,
which itself is unstable under renormalization-group flow. For $Q=4$, the flow structure is similar, except that the
tricritical point merges with the 2D critical fixed point~\cite{Nienhuis1979}, as shown schematically in
Fig.~\ref{f-10}(c).

Our numerical results suggest that the cases $k=1$ and $k=2$ follow the scenario shown in Fig.~\ref{f-10}(a). The
transition remains continuous for any $r$, and for finite $r$, the critical behavior always flows to the standard 2D
percolation fixed point. The mean-field fixed point at $1/r=0$ is therefore unstable and realized only in the singular
infinite-range limit.

The situation changes qualitatively for $k \ge 3$, as shown schematically in Fig.~\ref{f-10}(d). For sufficiently short
interaction ranges (large $1/r$), the system lies in the continuous-transition regime and again belongs to the
short-range 2D percolation universality class. Upon decreasing $1/r$, the system crosses into a first-order regime,
while the hybrid transition observed on CGs is unstable with respect to finite-range perturbations. The resulting
renormalization-flow structure is therefore qualitatively similar to the $Q=3$ or $Q=4$ Potts cases [Fig.~\ref{f-10}(b)
or (c)], depending on whether the boundary between the continuous and first-order regimes terminates at a tricritical
point or at a critical endpoint. Our present numerical results cannot distinguish these two possibilities.

For comparison, the numerical results reported in Ref.~\cite{Xue2024} suggest a different scenario, illustrated in
Fig.~\ref{f-10}(e). In that picture, the system exhibits three regimes as $r$ varies: a continuous-transition regime at
small $r$, a first-order regime at intermediate $r$, and a hybrid-transition regime at sufficiently large $r$,
continuously connected to the mean-field CG behavior. Such a scenario would imply that the mean-field hybrid fixed point
remains stable over a finite range of large but finite $r$. This would correspond to a finite-range hybrid critical
behavior distinct from both the CG hybrid behavior and the short-range 2D universality class. However, we do not find
numerical evidence supporting the
existence of such a stable hybrid regime. Instead, our results suggest that the apparent hybrid-like behavior observed
at very large but finite $r$ is more naturally interpreted as a finite-size crossover from the first-order regime toward
the mean-field CG limit.

\section{Discussion} \label{sec-con}

In this work, we systematically investigated the critical behavior of $k$-core percolation
in two dimensions with a finite interaction range $r$. By combining large-scale numerical simulations
with finite-size scaling analyses, we established a comprehensive phase diagram in
the $(k,r)$ plane, in which a continuous-transition regime is separated from a
first-order regime by a special boundary $(k_s,r_s)$. Below this boundary, the transition
is continuous and belongs to the universality class of standard 2D percolation, independent
of the interaction range $r$. Above it, the transition becomes discontinuous, with
no evidence of hybrid critical behavior.

From the renormalization-group perspective, the system remains 2D for any finite $r$.
Therefore, whenever the transition is continuous, its universality class is expected to
be controlled by the short-range fixed point rather than by the interaction range itself.
Our results further indicate that, throughout the continuous regime, the universality class
is also insensitive to $k$: the local constraint imposed by the $k$-core pruning process
is insufficient to destabilize the standard 2D percolation fixed point. In contrast, the
mean-field criticality associated with the CG limit is unstable with respect to finite-range
perturbations. For $k=1$ and $2$, any finite $r$ drives the system to the standard 2D percolation
universality class, whereas for $k \ge 3$, moving away from the singular point $1/r=0$
converts the mean-field hybrid transition into a first-order one.

Our findings also imply that the hybrid transition observed in infinite-dimensional systems
must emerge through a dimensional crossover as the spatial dimension increases. Understanding
this crossover mechanism remains an important theoretical challenge. In particular, key open
questions include whether there exists a finite upper critical dimension above which hybrid
transitions become stable, and how the crossover structure identified here evolves and
eventually connects to the mean-field scenario.

More broadly, these results clarify the mechanisms underlying continuous, first-order, and hybrid
transitions in constrained percolation models. We expect that the physical picture developed
here will be relevant to a wide range of systems involving cooperative constraints and spatial
correlations, including glassy dynamics, jamming phenomena, and the resilience of complex networks.

\section*{Acknowledgment}
The research was supported by the National Natural Science Foundation of China (under Grant No.
12275263) and the Quantum Science and Technology National Science
and Technology Major Project (under Grant No. 2021ZD0301900).

\appendix

\bibliography{ref.bib}

\end{document}